\documentclass[prd,twocolumn,aps,amsmath,amssymb,nofootinbib,preprintnumbers]{revtex4}

\usepackage{graphicx}
\usepackage{dcolumn}
\usepackage{bm}
\usepackage{amsmath}
\usepackage{amsfonts}

\newcommand{\beq}{\begin{eqnarray}}
\newcommand{\eeq}{\end{eqnarray}}

\usepackage{verbatim}

\newcommand{\p}{\partial}
\newcommand{\Ocal}{{\cal O}}

\newcommand{\dsigma}{\delta \sigma_1}
\newcommand{\dsigmad}{\delta \dot{\sigma}_1}

\newcommand{\Rcal}{{\cal R}}

\newcommand{\ixt}{\textrm{\tiny (2)}}
\newcommand{\largedot}{\mbox{\boldmath $\cdot$}}
\newcommand{\largeddot}{\mbox{\boldmath $\cdot \cdot$}}

\newcommand{\fnl}{f_\textrm{NL}}

\allowdisplaybreaks[1]

\begin{document}

\title{Comment on non-Gaussianity in hybrid inflation}
\author{Antti V\"aihk\"onen}
\affiliation{Department of Physical Sciences, University of Helsinki,\\
  and Helsinki Institute of Physics,\\ P.O. Box 64, FIN-00014 University of
  Helsinki, Finland}

\begin{abstract}
  In the literature there have been incompatible estimates for the amount of
  non-Gaussianity in hybrid inflation. In this note we point out the sources
  for the discrepancies and show that the results for the amount of
  non-Gaussianity in hybrid inflation obtained by two different methods,
  namely, perturbing Einstein equation to second order and the separate
  universe approach, indeed are compatible.  This provides confidence in the
  methods themselves and in the actual computation of non-Gaussianities.
\end{abstract}

\preprint{HIP-2005-26/TH}
\maketitle


{\it Introduction.} Recently there has been considerable interest in the
possible non-Gaussian component of the cosmological perturbations.
Non-Gaussian perturbations in hybrid inflation were first estimated using
consistent second order perturbation theory\footnote{This is during inflation.
  Complete second order perturbation theory connecting inflationary
  perturbations to the observed CMB anisotropies has so far not been
  developed.}  in \cite{Enqvist:2004bk}, where the formalism of Acquaviva et
al.~\cite{Acquaviva:2002ud} was used. The approach is to perturb the metric
and matter sides of the Einstein equation, and use the resulting equations to
obtain the dynamics of required quantities. These quantities are then used to
find out the second order curvature perturbation during inflation.

This method leads to a set of equations and, even after simplifying (although
motivated and not too constraining) assumptions, the final expression for the
second order curvature is complicated, containing several nonlocal and
time-integrated terms. An order of magnitude of the result was estimated
already in the original study \cite{Enqvist:2004bk}. It was later re-estimated
in \cite{Lyth:2005du}, where the result seemed to be much larger.

A completely different method, the $\delta N$ formalism or the separate
universe approach (see e.g.~\cite{Wands:2000dp}), for computing the second
order curvature perturbation in hybrid inflation was employed by Lyth and
Rodr{\'\i}guez in \cite{Lyth:2005fi}. The result disagrees with the earlier
estimates and implies an insignificant contribution from the transverse field,
$\sigma$, in hybrid inflation.

There is a discrepancy between the results obtained with the two different
methods, thus raising doubts on the validity of the methods.  However, we
argue that in both \cite{Enqvist:2004bk} and \cite{Lyth:2005du} the time
evolution of certain quantities is not properly taken into account. In this
brief comment we present a re-estimate of the original expression for the
second order curvature perturbation from the $\sigma$ field. We show that when
all the time evolutions are correctly taken into account the order of
magnitude estimate does indeed agree with the result obtained with the
separate universe approach. The second order curvature seems to be
proportional to the slow roll parameters. Such a small curvature alone would
make the resultant non-Gaussianity unobservable, but according to
\cite{Boubekeur:2005fj} there is a further suppression of the nonlinearity
parameter in this particular model due to the uncorrelated \cite{Lyth:2005du}
nature of the $\sigma$ non-Gaussianities.

{\it Original computation.} The original estimation of the non-Gaussianity was
obtained in \cite{Enqvist:2004bk} by extending the formalism of
\cite{Acquaviva:2002ud} for two scalar fields. The approach is to expand the
perturbations of the metric and the matter, which consists of two scalar
fields, up to second order. These perturbed quantities are then used to write
Einstein equations to second order. The curvature perturbation, defined for
one scalar field in the first order as \cite{Acquaviva:2002ud} $\Rcal = \psi +
H \frac{\delta \varphi}{\dot\varphi}$, is written for two scalar fields and
expanded to second order; here $\psi$ is metric perturbation, $H=\dot a /a$ is
the Hubble parameter, $\delta \varphi$ is the inflaton perturbation, and
$\dot\varphi$ is the time derivative of the background value of the inflaton
field. The Einstein equations are then used to obtain the evolution of $\Rcal$
to second order, which in turn is then used to estimate the amount of
non-Gaussianity.

The analytic calculation of the evolution equations in the case of two scalar
fields becomes complicated. To alleviate these difficulties two simplifying
assumptions are made in \cite{Enqvist:2004bk}, namely, it is assumed that the
background value $\sigma_0=0$ and that the potential does not have any terms
linear in $\sigma$. The latter assumption means that $\sigma_0=0$ indeed is a
local minimum.  The assumptions are well motivated and not too constraining,
and they clearly apply to many other models in addition to hybrid inflation,
whose potential is \cite{Enqvist:2004bk} $V = V_0 - \frac{m_0^2}{2} \sigma^2 +
\frac{\lambda}{4} \sigma^4 + \frac{m^2}{2} \varphi^2 + \frac{g^2}{2} \sigma^2
\varphi^2$. In fact, at least some number of e-folds after horizon exit of the
relevant scales, the form of the potential can be taken to be $V = V_0 +
\frac{m_\sigma^2}{2} \sigma^2 + \frac{m^2}{2} \varphi^2$, where $m_\sigma$ and
$m$ are the effective masses of $\sigma$ and $\varphi$, respectively.

The two constraints cause the second field, $\sigma$, to completely decouple
from the first order Einstein equations. Its behaviour in the first order is
only governed by the Klein-Gordon equation. At the second order, the
contribution of $\sigma$ becomes completely additive, i.e.~in the evolution
equations there are no $\varphi \sigma$-mixing terms. This enables one to make
use of the result of \cite{Acquaviva:2002ud} for the inflaton contribution in
\cite{Enqvist:2004bk}. Therefore, what is new in \cite{Enqvist:2004bk} is the
contribution coming from the transverse field $\sigma$.

The expression obtained in the original paper \cite{Enqvist:2004bk} for the
contribution of the $\sigma$ field perturbations to the second order curvature
perturbation is rather complicated, containing several nonlocal terms and
several time integrated terms. It reads
\begin{eqnarray} \label{eq:R2_orig}
  && \Rcal^{\ixt}_{\sigma} =  \frac{1}{\epsilon H M_P^2} \Big\{ \int
  \Big[ 6 H \Delta^{-1} \p_i ( \dsigmad \p^i
  \dsigma ) 
  \nonumber\\
  && + 4 \triangle^{-1} \p_i ( \dsigmad \p^i \dsigma )^{\largedot}
  - 2 ( \dsigmad )^2 + m^2_{\sigma} ( \dsigma)^2 
  \nonumber\\
  && + ( \epsilon - \eta) 6 H
  \Delta^{-2} \p_i (\p_k \p^k \dsigma \p^i \dsigma)^{\largedot} 
  \nonumber\\
  && + ( \epsilon - \eta ) H \Delta^{-2} \p_i \p^i (\p_k \dsigma \p^k
  \dsigma)^{\largedot} 
  \nonumber\\
  && - 3 \Delta^{-2} \p_i ( \p_k \p^k \dsigma \p^i
  \dsigma)^{\largeddot} 
  \nonumber\\
  && - \frac{1}{2} \Delta^{-2} \p_i \p^i ( \p_k \dsigma \p^k
  \dsigma)^{\largeddot} \Big] \,dt - \Delta^{-1} \p_i ( \dsigmad \p^i
  \dsigma)
  \nonumber\\
  && + 3 \Delta^{-2} \p_i ( \p_k \p^k \dsigma \p^i \dsigma)^{\largedot} +
  \frac{1}{2} \Delta^{-2} \p_i \p^i ( \p_k \dsigma \p^k \dsigma)^{\largedot}
  \nonumber\\
  && + 3 \epsilon H \Delta^{-2} \p_i ( \p_k \p^k \dsigma \p^i \dsigma)
  \nonumber\\
  && + \frac{\epsilon H}{2} \Delta^{-2} \p_i \p^i ( \p_k \dsigma \p^k \dsigma)
  \Big \}~,
\end{eqnarray}
where $\Delta^{-1}$ is the inverse Laplacian, $M_P \equiv (8 \pi G_N)^{-1/2}$
is the reduced Planck mass, $\dsigma$ is the first order perturbation of the
transverse scalar field $\sigma$, and $\epsilon \equiv \frac{M_P^2}{2} \big(
\frac{1}{V} \frac{\partial V}{\partial \varphi} \big)^2$ and $\eta \equiv
M_P^2 \frac{1}{V} \frac{\partial^2 V}{\partial \varphi^2}$ are slow roll
parameters; dot denotes derivative with respect to time.

In \cite{Enqvist:2004bk} the slow roll solution $\dsigma \propto
e^{-m^2_\sigma t / 3 H }$ was used to obtain an estimate for the time
derivative $|\dsigmad| \sim \frac{m^2_\sigma}{H} |\dsigma|$. (Double time
derivatives were estimated by $d^2/dt^2 \sim (m_\sigma^2 / H)^2$, but actually
one should use equation of motion to get rid of them). Since both fields,
$\varphi$ and $\sigma$, are effectively massless, the relation $|\dsigma| \sim
| \delta \varphi_1 |$ for the first order perturbations was used to
approximate
\begin{equation} \label{eq:approx1}
  \left| H \frac{\dsigma}{\dot{\varphi}_0} \right| \sim \left| H
  \frac{\delta \varphi_1}{\dot{\varphi}_0} \right| \equiv \left| \Rcal^{(1)}
  \right|~.
\end{equation}
In the first order Einstein equations $\sigma$ is completely decoupled, and
the situation is essentially that of a single field inflation. Therefore,
$\Rcal^{(1)}$ stays constant outside horizon.

Only order of magnitude estimate was pursued and cancelling spatial derivative
operators were neglected, e.g.~$| \Delta^{-1} \partial_i \Rcal^{(1)}
\partial^i \Rcal^{(1)} | \sim | \Rcal^{(1)} |^2$. For the estimation of the
time integral the quantities $H$, $\epsilon$, $\eta$, $m_{\sigma}$ and
$\dsigma$ in the integrand were taken to be constants. The original result in
\cite{Enqvist:2004bk} for the estimate of the second order perturbation due to
$\sigma$ reads
\begin{equation} \label{eq:result_orig}
  \Rcal_\sigma^{(2)} \sim \Ocal(\epsilon, \eta, \frac{m_\sigma^2}{H^2} ) \;
  | \Rcal^{(1)} |^2~.
\end{equation}
Since $\eta_\sigma \equiv M_P^2 \frac{1}{V} \frac{\partial^2 V}{\partial
  \sigma^2} \simeq \frac{m_\sigma^2}{H^2}$, the entire coefficient is of the
order slow roll parameters.

{\it Re-estimate by Lyth and Rodr{\'\i}guez.}  Later Lyth and Rodr{\'\i}guez
\cite{Lyth:2005du} made a re-estimation of Eq.~(\ref{eq:R2_orig}) by inserting
initial conditions and writing the equation as a definite integral.  They,
however, used the same estimates, Eq.~(\ref{eq:approx1}) (with $\Rcal^{(1)}
\sim \mbox{const}$) and $|\dsigmad| \sim \frac{m^2_\sigma}{H} |\dsigma|$, as
the original study \cite{Enqvist:2004bk}.  Similarly, they also assumed $H$,
$m_\sigma$, and $\epsilon$ to be constants, ending up with\footnote{They
used a different definition for the second order curvature, but that is
irrelevant here.}  \cite{Lyth:2005du}
\begin{eqnarray}
  && \Rcal^{(2)}(t) - \Rcal^{(2)}(t_i) = \frac{1}{\epsilon H M_P^2}
  \int^t_{t_i} \Big[ 6 H \Delta^{-1} \p_i ( \dsigmad \partial^i \dsigma )
  \nonumber \\
  && + m^2_{\sigma} ( \dsigma )^2 - 2 ( \dsigmad )^2 \Big] dt
  \; \sim \; \Delta N \frac{m_\sigma^2}{H^2} | \Rcal^{(1)} |^2~.
\end{eqnarray}

Since $\Delta N$ is the number of e-folds, which can very well be $\sim 60$,
this result implies a much larger $\sigma$ contribution to the second order
curvature perturbation than in the original study.

{\it Separate universe approach.} In addition to the cosmological perturbation
theory approach \cite{Enqvist:2004bk}, there also exists a recent computation
\cite{Lyth:2005fi} of the second order curvature perturbation in hybrid
inflation using the separate universe approach.

The general idea of the separate universe approach (see
e.g.~\cite{Wands:2000dp} for a concise description) is to consider each point
in space as being surrounded by a homogeneous FRW universe. Each point then
has its own expansion parameter $N$, i.e~local number of e-folds, independent
of the value of the expansion parameter (or any quantity) in other points.
This expansion parameter depends on the values of relevant quantities, such as
unperturbed scalar fields, at that point. The complete, inhomogeneous,
behaviour of the universe is obtained when all the separately treated points
are patched together.

The curvature perturbation, $\zeta$, is in \cite{Lyth:2005fi} defined by
\begin{equation}
  g_{ij} = a^2 (t) e^{2 \zeta (t,\boldsymbol x)} \gamma_{ij} (t,\boldsymbol
  x)~,
\end{equation}
where, within inflationary context, $\gamma_{ij}$ contains the tensor
perturbation which we do not consider here, (see \cite{Lyth:2005du} for more
details). Up to second order in scalar field perturbations ($\delta \phi_i
\equiv \delta \phi_i (t,\boldsymbol x)$) $\zeta$ is obtained from
\cite{Lyth:2005fi}
\begin{equation}
  \zeta (t,\boldsymbol x) = \sum_i N_{,i} (t) \delta \phi_i + \frac{1}{2}
  \sum_{ij} N_{,ij} (t) \delta \phi_i \delta \phi_j~,
\end{equation}
where $N_{,i} \equiv \frac{\partial N}{\partial \phi_i}$ and $N_{,ij} \equiv
\frac{\partial^2 N}{\partial \phi_i \partial \phi_j}$.

Adapting the notation $V = V_0 (1 + \frac{1}{2} \eta \varphi^2 + \frac{1}{2}
\eta_\sigma \sigma^2 )$ for the potential, the curvature perturbation reads
\cite{Lyth:2005fi}
\begin{equation} \label{eq:zeta_LR}
  \zeta = \frac{\delta \varphi}{\eta \varphi} - \frac{\eta}{2} \Big(
  \frac{\delta \varphi}{\eta \varphi} \Big)^2 + \frac{\eta_\sigma}{2} e^{2
  \Delta N (\eta - \eta_\sigma)} \Big( \frac{\delta \sigma}{\eta \varphi}
  \Big)^2~.
\end{equation}
The fields $\varphi$ and $\sigma$ are assumed to be massless and their
perturbations, $\delta \varphi$ and $\delta \sigma$, are assumed to have the
same spectrum $\big( \frac{H_*}{2 \pi} \big)^2$ in \cite{Lyth:2005fi}.
Therefore, we can set $|\zeta_1| \sim \big| \frac{\delta \varphi}{\eta
  \varphi} \big| \sim \big| \frac{\delta \sigma}{\eta \varphi} \big|$ and
estimate the last term in Eq.~(\ref{eq:zeta_LR}), i.e.~the contribution of
$\sigma$ to the second order curvature perturbation as
\begin{equation} \label{eq:R2_estimate_LR}
  \zeta_{2,\sigma} \sim \Ocal(\eta_\sigma)\; e^{2 \Delta N (\eta -
  \eta_\sigma)} \; |\zeta_1|^2~.
\end{equation}

{\it Source for the discrepancy.}  The seeming discrepancy between the two
different methods basically comes down to articles \cite{Enqvist:2004bk} and
\cite{Lyth:2005du}, or more precisely, assuming $\epsilon$ to be constant, and
to the assumption in Eq.~(\ref{eq:approx1}), i.e.
\begin{equation} \label{eq:delta_equal}
  | \dsigma (t) | \sim | \delta \varphi_1 (t) |~,
\end{equation}
which is not generally justified. Eq.~(\ref{eq:delta_equal}) only holds
immediately after horizon exit ($t=t_i$), when the amplitude of the
perturbation of any effectively massless field $f$ is $|\delta f| \sim H$.

Using the slow roll equations (outside horizon) we obtain
\begin{eqnarray} \label{eq:time_evol}
  \dsigma(t) &=& \dsigma(t_i) e^{-\eta_\sigma \Delta N}~,
  \nonumber \\
  \delta \varphi_1 (t) &=& \delta \varphi_1 (t_i) e^{-\eta_\varphi \Delta N}~,
  \\ 
  \varphi_0 (t) &=& \varphi_0 (t_i) e^{-\eta_\varphi \Delta N}~,
  \nonumber
\end{eqnarray}
where $\Delta N = H \Delta t$ is the number of e-folds since $t_i$; thus, we
obtain $\dsigmad = - \eta_\sigma H \dsigma$, $\delta \dot{\varphi}_1 =
-\eta_\varphi H \delta \varphi_1$, and $\dot{\varphi}_0 = -\eta_\varphi H
\varphi_0$.  Since $\epsilon \sim {\dot{\varphi}}_0^2 / H^2 M_P^2$ we can also
readily write
\begin{equation} \label{eq:time_evol_epsilon}
  \epsilon (t) = \epsilon_i \, e^{- 2 \eta_\varphi \Delta N}~,
\end{equation}
where we have denoted $\epsilon_i \equiv \epsilon (t_i)$.

Now, it is immediately clear that $| \Rcal^{(1)} | = | H \delta \varphi_1 /
\dot{\varphi}_0 |$ stays constant, but one also sees that $\Delta N$ e-folds
after horizon exit
\begin{equation}
  | H \frac{\dsigma}{\dot{\varphi}_0} | \sim | \Rcal^{(1)} | \; |
    \frac{\dsigma}{\delta \varphi_1} | \sim e^{\Delta N \, (\eta_\varphi -
    \eta_\sigma)} | \Rcal^{(1)} |~.
\end{equation}
For the estimation of the time integral in Eq.~(\ref{eq:R2_orig}) the
important point is that one may not move $1/\epsilon$ and $\dsigma$ into and
out of the time integral.

{\it Re-estimate of non-Gaussianity.} Now we present a re-estimate of
Eq.~(\ref{eq:R2_orig}) using the time evolutions expressed in
Eqs.~(\ref{eq:time_evol}) and (\ref{eq:time_evol_epsilon}). First, we notice
that
\begin{eqnarray} \label{eq:reduction}
  && 6 H \Delta^{-1} \partial_i ( \dsigmad \partial^i \dsigma ) + 2
  \Delta^{-1} \partial_i ( \dsigmad \partial^i \dsigma )^{\largedot}
  \nonumber \\
  && - ( \dsigmad )^2 + m_\sigma^2 ( \dsigma )^2
  \nonumber \\
  && = 2 \Delta^{-1} \partial_i \big[ ( 3 H \dsigmad + \delta \ddot{\sigma}_1
  + m_\sigma^2 \dsigma ) \partial^i \dsigma \big ]
  \nonumber \\
  && = 0~,
\end{eqnarray}
since $3 H \dsigmad + \delta \ddot{\sigma}_1 + m_\sigma^2 \dsigma = 0$ outside
horizon.  Thus, Eq.~(\ref{eq:R2_orig}) can be written\footnote{We use the
  indefinite integral here, instead of definite one with initial conditions,
  since the initial second order curvature, or at least initial
  non-Gaussianity, is supposedly small enough to be safely neglected
  \cite{Lyth:2005du}.}
\begin{eqnarray} \label{eq:R2_final}
  \Rcal^{\ixt}_{\sigma} &=&  \frac{1}{\epsilon H M_P^2} \Big\{ \int^t
  \Big[ 2 \triangle^{-1} \p_i ( \dsigmad \p^i \dsigma )^{\largedot}
  - ( \dsigmad )^2
  \nonumber\\
  && + 2 H (\epsilon - \eta ) \dot{\gamma}_\sigma - \ddot{\gamma}_\sigma \Big]
  dt - \Delta^{-1} \p_i ( \dsigmad \p^i \dsigma)
  \nonumber\\
  && + \dot{\gamma}_\sigma + \epsilon H \gamma_\sigma \Big\}
  \nonumber\\
  &=& \frac{1}{\epsilon H M_P^2} \Big\{ \int^t \Big[
  - ( \dsigmad )^2 + 2 H \epsilon \dot{\gamma}_\sigma \Big] dt
  \nonumber\\
  && + \Delta^{-1} \p_i ( \dsigmad \p^i \dsigma) + H (\epsilon - 2 \eta)
  \gamma_\sigma \Big\}~,
\end{eqnarray}
where we have denoted
\begin{eqnarray}
  \gamma_\sigma & \equiv & 3 \Delta^{-2} \p_i ( \p_k \p^k \dsigma \p^i
  \dsigma)
  \nonumber\\
  &+& \frac{1}{2} \Delta^{-2} \p_i \p^i ( \p_k \dsigma \p^k \dsigma)~.
\end{eqnarray}
The last step is due to time derivatives within the time integral ($\eta$ and
$H$ are assumed to be constants).

Eq.~(\ref{eq:reduction}) is used to get rid of the first term in
Eq.~(\ref{eq:R2_orig}), $\frac{1}{\epsilon H M_P^2} \int 6 H \Delta^{-1} \p_i
( \dsigmad \p^i \dsigma ) dt$, which would give too large a contribution to
the estimate. Note that the order of magnitude estimate does not take into
account possible cancellations and, therefore, provides only an upper limit.
However, the cancellations can be treated explicitly, as is done here.

For estimation purposes we also adopt the potential used in
\cite{Lyth:2005fi}, $V = V_0 (1 + \frac{1}{2} \eta \varphi^2 + \frac{1}{2}
\eta_\sigma \sigma^2 )$. The slow roll parameters $\eta$, and $\eta_\sigma$
are constants, and we also set $H$ to be constant.  The time evolutions of
$\dsigma$, $\delta \varphi_1$, $\varphi_0$, and $\epsilon$ are given by
Eqs.~(\ref{eq:time_evol}) and (\ref{eq:time_evol_epsilon}).

Since the order of magnitude estimate anyway gives an upper limit, and since
$\epsilon \leq \epsilon_i$ for any time $t \geq t_i$, we replace $\epsilon$
with $\epsilon_i$ except in the factor $\frac{1}{\epsilon}$. We again neglect
cancelling orders of spatial derivative operators\footnote{Because of this,
  both terms in $\gamma_\sigma$ are effectively the same.}, and put $\dsigmad
= - \eta_\sigma H \dsigma$. The estimate for the second order curvature
perturbation, Eq.~(\ref{eq:R2_final}), thus becomes
\begin{eqnarray}
  \Rcal^{\ixt}_{\sigma} &\sim& \frac{1}{\epsilon H M_P^2} \Big\{ \int^t \Big[
  \eta_\sigma^2 H^2 |\dsigma|^2 + \epsilon_i \eta_\sigma H^2 |\dsigma|^2
  \Big] dt
  \nonumber\\
  && + \eta_\sigma H |\dsigma|^2 + \epsilon_i H |\dsigma|^2 + \eta H
  |\dsigma|^2 \Big\}
  \nonumber\\
  &\sim& \frac{1}{\epsilon H M_P^2} \Big\{ \Ocal(\epsilon_i,\eta_\sigma)
  \int^t \eta_\sigma H^2 |\dsigma|^2 \; dt
  \nonumber\\
  && + \Ocal(\epsilon_i,\eta,\eta_\sigma)\; H |\dsigma|^2 \Big\}~.
\end{eqnarray}
We have now two terms to evaluate, namely
\begin{eqnarray}
  && \frac{1}{\epsilon H M_P^2} \; H |\dsigma|^2 = \frac{|\dsigma
  (t_i)|^2}{\epsilon_i M_P^2} e^{2 \Delta N (\eta - \eta_\sigma)} 
  \nonumber\\
  && = \Big| H \frac{\dsigma (t_i)}{\dot{\varphi}_0 (t_i)} \Big|^2 e^{2 \Delta
  N (\eta - \eta_\sigma)}
  \nonumber\\
  && = e^{2 \Delta N (\eta - \eta_\sigma)} |\Rcal^{(1)}|^2~,
\end{eqnarray}
and
\begin{eqnarray}
  && \frac{1}{\epsilon H M_P^2} \int^t H^2 \eta_\sigma |\dsigma|^2 \; dt
  \nonumber\\
  &&  = \frac{|\dsigma (t_i)|^2}{\epsilon M_P^2} \eta_\sigma \int^{\Delta N}
  e^{-2  \eta_\sigma N} dN
  \nonumber\\
  && \sim \frac{|\dsigma (t_i)|^2}{\epsilon M_P^2} e^{-2 \eta_\sigma \Delta N}
  = e^{2 \Delta N (\eta - \eta_\sigma)} |\Rcal^{(1)}|^2~.
\end{eqnarray}

Therefore, our final estimate reads
\begin{equation} \label{eq:final_estimate}
  \Rcal^{\ixt}_{\sigma} \sim \Ocal(\epsilon_i,\eta,\eta_\sigma) \; e^{2 \Delta
  N (\eta - \eta_\sigma)} \; |\Rcal^{(1)}|^2~.
\end{equation}
Comparing to the estimate in Eq.~(\ref{eq:R2_estimate_LR}) one sees that the
two results are of the same form and of the same order. Instead of
$\Ocal(\epsilon_i,\eta,\eta_\sigma)$ Eq.~(\ref{eq:R2_estimate_LR}) only has a
factor $\Ocal(\eta_\sigma)$. Here we have, however, provided only an order of
magnitude estimate and there still is a possibility for cancellations which
may have been overlooked. It may be worth pointing out that if the nonlocal
terms are discarded in Eq.~(\ref{eq:R2_final}) and only the first term is
estimated, we obtain an $\Ocal(\eta_\sigma)$ coefficient only.

{\it Discussion.} The maximum number of e-folds for the observable scales is
$\sim 60$ and observational limits for the spectral index of the curvature
perturbation require $|\eta| \lesssim 0.01$. It is therefore unlikely that the
exponential factor $e^{2 \Delta N (\eta - \eta_\sigma)}$ would provide any
significant enhancement, and the overall factor of the order slow roll
parameters gives the magnitude of the result.

The quantity measured in CMB experiments is the nonlinearity parameter $\fnl$,
which is related to $\Rcal^{(2)}$, but according to \cite{Boubekeur:2005fj} it
is highly suppressed in this scenario. The projected sensitivity (using first
order perturbation theory) for an ideal experiment is no better than $\fnl
\sim 1$ even including polarisation \cite{Babich:2004yc}. It was later
realized by Vernizzi \cite{Vernizzi:2004nc} that the the second order
curvature perturbation defined in \cite{Acquaviva:2002ud} has an artificial
time evolution $\dot{\Rcal}^{(2)} \sim (2 \dot\epsilon - \dot\eta)
{\Rcal^{(1)}}^2$. However, even taking this small effect into account, it
seems safe to say that the non-Gaussianity produced in the hybrid scenario by
the $\sigma$ field seems to be too small to be observed unless some key
aspects of the scenario are changed.

There are still conceptual and practical problems with computing
non-Gaussianities. The second order theory is not yet well established and the
connection between theoretical calculations and CMB observations is far from
complete. One problem is that the usual scalar-vector-tensor decomposition of
the perturbations is inherently non-local \cite{Ellis:1990gia}. These
non-localities do not appear in the first order, but in the second order
equations there are terms like $\Delta^{-1}(\partial_i g \partial^i g )$ and
$\Delta^{-1}(g \Delta g )$, where $g$ represents a generic perturbation. The
physical interpretation of these terms is not clear.

Lyth and Rodr{\'\i}guez apply the separate universe approach in
\cite{Lyth:2005fi} to compute the second order perturbations, or
non-Gaussianity, in various scenarios including hybrid inflation. The
formalism they use does not produce non-local terms involving the inverse
Laplacian $\Delta^{-1}$, and they state that \cite{Lyth:2005fi} ``\ldots such
terms must cancel if correctly evaluated.''  Needless to say, this is quite a
strong claim. It is actually not certain whether the separate universe
approach is completely correct when expanded to second order in perturbations.
Indeed, in \cite{Wands:2000dp} it was stated that any nonlinear interaction
introduces mode-mode couplings which undermine the separate universe picture.
As an example, recent studies of non-Gaussianities in preheating
\cite{Enqvist:2004ey} demonstrate that these mode-mode couplings seem to be
important.

Despite all the problems and ambiguities in the different second order
formalisms it is comforting that the two different approaches discussed here
can now be seen to produce the same result for the second order curvature
perturbation and, therefore, for the amount of non-Gaussianity in hybrid
inflation.

{\it Acknowledgements.} The author would like to thank Kari Enqvist and Asko
Jokinen for useful discussions and commenting on the manuscript and Filippo
Vernizzi for enlightening discussions. This work is supported by the Magnus
Ehrnrooth Foundation.





\begin{thebibliography}{9}

\bibitem{Enqvist:2004bk}
  K.~Enqvist and A.~Vaihkonen,
  JCAP {\bf 0409} (2004) 006
  [arXiv:hep-ph/0405103].

\bibitem{Acquaviva:2002ud}
  V.~Acquaviva, N.~Bartolo, S.~Matarrese and A.~Riotto,
  Nucl.\ Phys.\ B {\bf 667} (2003) 119
  [arXiv:astro-ph/0209156].

\bibitem{Lyth:2005du}
  D.~H.~Lyth and Y.~Rodriguez,
  arXiv:astro-ph/0502578.

\bibitem{Wands:2000dp}
  D.~Wands, K.~A.~Malik, D.~H.~Lyth and A.~R.~Liddle,
  Phys.\ Rev.\ D {\bf 62} (2000) 043527
  [arXiv:astro-ph/0003278].

\bibitem{Lyth:2005fi}
  D.~H.~Lyth and Y.~Rodriguez,
  arXiv:astro-ph/0504045.

\bibitem{Boubekeur:2005fj}
  L.~Boubekeur and D.~H.~Lyth,
  arXiv:astro-ph/0504046.

\bibitem{Babich:2004yc}
  D.~Babich and M.~Zaldarriaga,
  Phys.\ Rev.\ D {\bf 70} (2004) 083005
  [arXiv:astro-ph/0408455].

\bibitem{Vernizzi:2004nc}
  F.~Vernizzi,
  Phys.\ Rev.\ D {\bf 71} (2005) 061301
  [arXiv:astro-ph/0411463].

\bibitem{Ellis:1990gia}
  G.~F.~R.~Ellis, M.~Bruni and J.~Hwang,
  Phys.\ Rev.\ D {\bf 42} (1990) 1035.

\bibitem{Enqvist:2004ey}
  K.~Enqvist, A.~Jokinen, A.~Mazumdar, T.~Multamaki and A.~Vaihkonen,
  Phys.\ Rev.\ Lett.\  {\bf 94} (2005) 161301
  [arXiv:astro-ph/0411394].

\end{thebibliography}
\end{document}